\documentclass[aps, reprint, superscriptaddress, prl, 10pt]{revtex4-1}
\usepackage{graphicx}
\usepackage{dcolumn}
\usepackage{bm}
\usepackage{amsmath}
\usepackage[latin1]{inputenc}
\usepackage{nicefrac}
\usepackage{color}

\begin{document}

\title{Quantum spin chain as a potential realization of the Nersesyan-Tsvelik model}

\author{C. Balz}
\author{B. Lake}
\affiliation{Helmholtz-Zentrum Berlin für Materialien und Energie, 14109 Berlin, Germany}
\affiliation{Institut für Festkörperphysik, Technische Universität Berlin, 10623 Berlin, Germany}
\author{H. Luetkens}
\author{C. Baines}
\affiliation{Laboratory for Muon-Spin Spectroscopy, Paul Scherrer Institut, 5232 Villigen, Switzerland}
\author{T. Guidi}
\affiliation{ISIS Facility, STFC Rutherford Appleton Laboratory, Oxfordshire OX11 0QX, UK}
\author{M. Abdel-Hafiez}
\affiliation{Leibniz Institute for Solid State and Materials Research, IFW, 01171 Dresden, Germany}
\affiliation{Physics department, Faculty of science, Fayoum University, 63514 Fayoum, Egypt}
\author{A.U.B. Wolter}
\author{B. Büchner}
\affiliation{Leibniz Institute for Solid State and Materials Research, IFW, 01171 Dresden, Germany}
\author{I.V. Morozov}
\author{E.B. Deeva}
\affiliation{Faculty of Chemistry, Lomonosov Moscow State University, 119991 Moscow, Russia}
\author{O.S. Volkova}
\author{A.N. Vasiliev}
\affiliation{Faculty of Physics, Lomonosov Moscow State University, 119991 Moscow, Russia}
\affiliation{Theoretical Physics and Applied Mathematics Department, Ural Federal University, 620002 Ekaterinburg, Russia}

\date{\today}

\begin{abstract}
It is well established that long-range magnetic order is suppressed in magnetic systems whose interactions are low-dimensional. The prototypical example is the S-\nicefrac{1}{2} Heisenberg antiferromagnetic chain (S-\nicefrac{1}{2} HAFC) whose ground state is quantum critical. In real S-\nicefrac{1}{2} HAFC compounds interchain coupling induces long-range magnetic order although with a suppressed ordered moment and reduced N\'{e}el temperature compared to the Curie-Weiss temperature. Recently, it was suggested that order can also be suppressed if the interchain interactions are frustrated, as for the Nersesyan-Tsvelik model. Here, we study the new S-\nicefrac{1}{2} HAFC, (NO)[Cu(NO$_3$)$_3$]. This material shows extreme suppression of order which furthermore is incommensurate revealing the presence of frustration consistent with the Nersesyan-Tsvelik model.
\end{abstract}

\maketitle


The spin-\nicefrac{1}{2} Heisenberg chain with antiferromagnetic interactions $J$, is an archetypal model of condensed matter physics. This system lies at the Luttinger Liquid Quantum critical point where quantum critical fluctuations destroy long-range magnetic order and give rise to algebraically decaying correlations \cite{Sch86,Lak05}. The fundamental excitations are spinons, which unlike conventional spin-waves, possess a fractional spin quantum number (S-\nicefrac{1}{2}) \cite{fad81}. The dynamical structure factor of the S-\nicefrac{1}{2} HAFC has been tackled using several theoretical approaches \cite{Mue81,Kar97}. Major advances occurred in 2005 when a near-exact calculation was published \cite{Cau05,Cau05_2,Cau06,Bar13} that has been verified experimentally \cite{Lak13,Mou13}.

\begin{figure}
\includegraphics[width=\columnwidth]{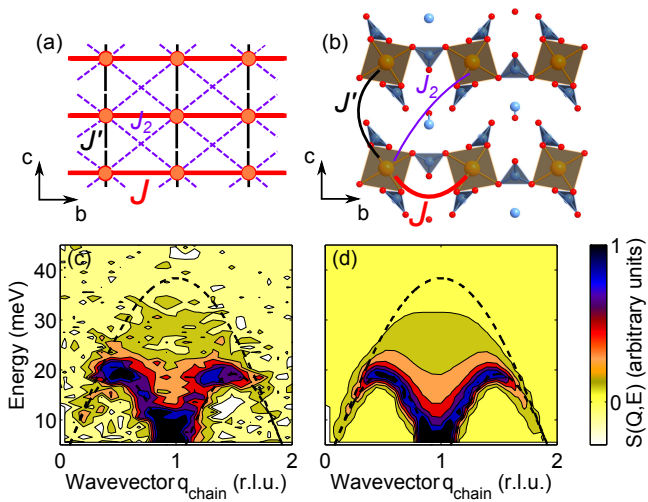}
\caption{(a) Schematic of the Nersesyan-Tsvelik model. (b) Crystal structure: orange plaquettes represent CuO$_4$, blue triangles NO$_3^-$ and  dumbbells NO$^+$. (c) Background-subtracted INS data at 5.5\,K showing the spinon continuum as a function of wavevector along the chain and energy. The data are integrated over the perpendicular wavevectors. The background was modeled by interpolating the intensity at even-integer $q_{\text{chain}}$ where the magnetic signal is weak. (d) Simulation of theoretical spinon continuum \cite{Cau06} with $J=142$\,K. The black dashed lines denote the upper and lower boundary of the 2-spinon continuum.}\label{Fig:merlin}
\end{figure}

An ideal S-\nicefrac{1}{2} HAFC cannot exist in real materials because even an infinitesimal interchain coupling gives rise to long-range magnetic order at suppressed but finite temperatures. The interchain coupling also effects the excitation spectrum and the spinons become confined resulting in spin-waves and a longitudinal mode at low energies \cite{Ten95,Lak00,Lak05_2,Lak05_3}. It is well established that frustration suppresses long-range magnetic order and can often drive it incommensurate, indeed incommensurate magnetism is a strong signature of frustrated interactions. A theoretically well-studied example is the "Nersesyan-Tsvelik" model \cite{Ner03} where S-\nicefrac{1}{2} HAFCs are coupled in a plane by a combination of frustrated antiferromagnetic nearest-neighbor ($J'$) and next-nearest neighbor ($J_2$) interactions (Fig.\,\ref{Fig:merlin}a). At the special point $J'=2J_2$, these interactions effectively cancel completely suppressing long-range order and deconfining the spinons. The ground state has variously been predicted to be a resonating valence bond \cite{Ner03,Tsv04}, a valence bond crystal \cite{Sta04} or a gapless spin liquid \cite{Mou04}. In this paper we investigate the S-\nicefrac{1}{2} HAFC (NO)[Cu(NO$_3$)$_3$], which is a potential realization of the Nersesyan-Tsvelik model. 

The crystal structure of (NO)[Cu(NO$_3$)$_3$] (Fig.\,\ref{Fig:merlin}b) suggests that the spin-\nicefrac{1}{2} Cu$^{2+}$ ions are coupled by strong antiferromagnet interactions $J$ along the $b$ axis and these chains are weakly coupled in the $bc$ plane by $J'$ and $J_2$. Inspection of the exchange paths suggests that $J' \approx 2J_2$ while the interlayer coupling is weak \cite{Vol10}. The susceptibility is typical of one-dimensional magnetism, with a broad maximum at $\sim$100\,K. Assuming $J'=2J_2$, fitting gives best agreement for $J=170$\,K and $-0.05<J'/J<0.09$ \cite{Vol10}. ESR measurements down to 3.4\,K show no signs of short- or long-range magnetic order. Band structure calculations support the picture of a S-\nicefrac{1}{2} HAFC with intrachain coupling $J\approx230$\,K \cite{Jan10}. These calculations however suggest weak but unfrustrated interchain couplings $J'<3$\,K and $J_2\ll J'$. Raman scattering shows a broad magnetic continuum due to free spinon excitations and yields $J\approx150$\,K \cite{Gne12}. Suppression of the magnetic scattering below $T^*\approx100$\,K is attributed to the dynamical interplay of spin and lattice degrees of freedom.

Here we investigate (NO)[Cu(NO$_3$)$_3$] using inelastic neutron scattering (INS), heat capacity ($C_p$) and muon spin relaxation ($\mu$SR). The INS at 5.5\,K shows spinon excitations typical of a S-\nicefrac{1}{2} HAFC with $J=142(3)$\,K, whereas the $C_p$ and $\mu$SR reveal a transition to long-range magnetic order at the highly suppressed N\'{e}el temperature of $T_{N}=585(5)$\,mK. The magnetic order is an incommensurate spin density wave implying the presence of frustrated interchain coupling consistent with the Nersesyan-Tsvelik model.


Single crystals of (NO)[Cu(NO$_3$)$_3$] were grown from a solution of metallic copper dissolved in water-free HNO$_3$ enriched with N$_2$O$_4$ and N$_2$O$_5$. This mixture was sealed in glass ampoules and cooled from 80 to 0°C over two weeks. Plate-like single crystals formed with dimensions $(0.5-2)\times(2-5)\times(5-10)$\,mm$^3$. They were dried and their phase purity was verified by X-ray powder diffraction \cite{Zna04}. The crystals are highly hygroscopic and contact with air was prevented.

The INS experiment was performed on the time-of-flight spectrometer MERLIN at the ISIS Facility, Rutherford Appleton Laboratory, UK. Two single crystals (total mass 216\,mg) were coated with \textsc{cytop} to protect them from air and were co-aligned with the $ab$ plane horizontal. An incident energy of 56.1\,meV and chopper speed of 200\,Hz were used giving an elastic resolution of 3.5\,meV. Measurements took place at 5.5\,K.

Low temperature specific heat measurements were performed by a Quantum Design Physical Properties Measurement System down to 0.4\,K and up to 9\,T using a relaxation technique. The raw data shows a broad peak centered at 1.82\,K caused by a copper nitrate trihydrate (Cu(NO$_3$)$_2\cdot$2.5H$_2$O) impurity. This impurity phase was $\sim$8\% and its heat capacity was fitted to the isolated S-\nicefrac{1}{2} dimer-model and subtracted from the data (see Supplemetal Material \cite{Sup14}).

The $\mu$SR measurements took place on the LTF spectrometer at the Paul Scherrer Institut, Switzerland. Five single crystals (total mass 133\,mg) were co-aligned on a silver cell with their $c$ axes parallel to the muon beam and their $b$ axes vertical. Zero-, longitudinal- and transverse-field measurements were performed for temperatures from 19 to 800\,mK. The data were analyzed using the \textsc{musrfit} software package \cite{Sut12}.


\begin{figure}
\includegraphics[width=0.85\columnwidth]{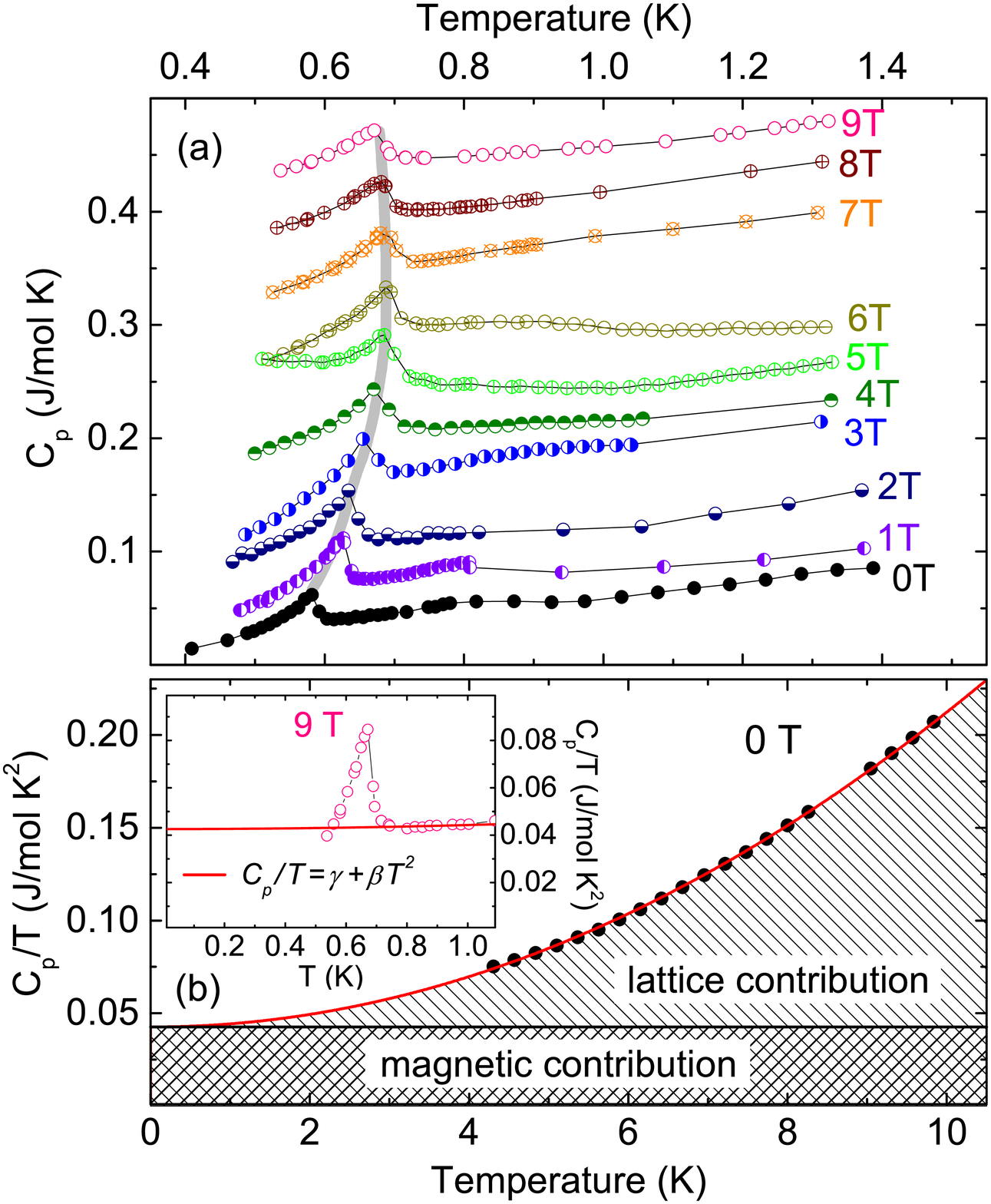}
\caption{(a) Temperature dependence of the specific heat at various magnetic fields. The measurements are shifted vertically with respect to each other. The gray line denotes the phase boundary. (b) $C_p/T$ vs. $T$ for $B=0$\,T (inset: $B=9$\,T). Regions contaminated by (Cu(NO$_3$)$_2\cdot$2.5H$_2$O) are excluded. A fit to $C_p/T=\gamma+\beta T^2$ is represented by the red line in (b) and in the inset with $\gamma=0.043$\,J/mol\,K$^2$ and $\beta=0.0017$\,J/mol\,K$^4$.}\label{Fig:specific_heat}
\end{figure}

The INS data at 5.5\,K is presented in Fig.\,\ref{Fig:merlin}c as a function of wavevector along the chain and energy. The clear presence of a spinon continuum proves that (NO)[Cu(NO$_3$)$_3$] is highly one-dimensional. The excitations are gapless at integer values of $q_{\text{chain}}$ showing that the magnetic structure is commensurate along the chain and that the interactions do not compete in this direction. The excitation spectrum was compared to the theoretical solution of J.S. Caux {\em et al.} for the dynamical structure factor of the S-\nicefrac{1}{2} HAFC at $T=0$\,K \cite{Cau06}. The theoretical scattering was convolved with the instrumental resolution and the mosaic spread of the sample (Fig.\,\ref{Fig:merlin}d). From the sharp onset of the excitations at $q_{\text{chain}}=0.5$ which occurs theoretically at $E=\pi J/2$, the intrachain coupling was found to be $J=142(3)$\,K. This value is similar although smaller than earlier estimations of $J=170$\,K from susceptibility data \cite{Vol10}, $J\approx230$\,K from band structure calculations \cite{Jan10}, and $J\approx150$\,K from Raman scattering \cite{Gne12}. INS, however, gives the most reliable value because of its sensitivity to the magnetic coupling.

The background subtracted specific heat is shown in Fig.\,\ref{Fig:specific_heat}. At $B=0$\,T a phase transition to long-range magnetic order takes place at $T_N=0.58(1)$\,K. $T_N$ is very small compared to $J$ suggesting that the ordered moment, $m$, in the ground state is also suppressed. Integrating $C(T)/T$ from 0 to 0.58\,K gives an upper estimate for the entropy released at $T_N$ of $S_{mag}=0.020$\,J/mol\,K which is only 0.34\,\% of the total entropy ($S=R\ln2$). An extensive search using single crystal neutron diffraction failed to find the magnetic Bragg peaks. This null result gives an upper limit of $m\leq0.01\,\mu_B$, which is highly reduced from the fully ordered value of $1\,\mu_B/\text{Cu}^{2+}$.

Above $T_N$ the specific heat is expected to take the form $C=\gamma T+\beta T^3$ where the linear term is the Luttinger Liquid contribution for the S-\nicefrac{1}{2} HAFC and the cubic term is due to phonons \cite{Voi95}. The small contribution from nuclear spins ($<$10\%) is ignored. Fitting this expression to the data (red line in Fig.\,\ref{Fig:specific_heat}b) gives $\gamma=0.043(1)$\,J/mol\,K$^2$ thus the magnetic exchange interaction $J$ can be estimated via the expression $\gamma=2N_{\text{A}}k_{\text{B}}^2/3J$ and gives $J=130(3)$\,K \cite{Bon64}. This value agrees well with $J=142(3)$\,K from INS. Magnetic fields up to 6\,T produce a small increase in $T_N$ in contrast to the field-induced reduction of $T_N$ in conventional antiferromagnets \cite{Jon01}. This suggests that in (NO)[Cu(NO$_3$)$_3$] a field suppresses the quantum fluctuations due to low-dimensionality or frustration that destabilize the order.

The suppression of the ordering temperature compared to the intrachain interaction $J$ represented by $f=|J|/T_N=243$, can be compared to other S-\nicefrac{1}{2} HAFCs. The most extensively studied spin chain, KCuF$_3$, has a relatively large interchain coupling ($J'/J\approx0.01$) yielding an ordered moment of $m=0.49(7)\,\mu_B$ and $f=390\text{K}/39\text{K}=10$ \cite{Hut69,Lak13}. A more 1D example is Sr$_2$CuO$_3$ with a very small interchain interaction ($J'/J\approx10^{-5}$), giving an ordered moment of only $m=0.06(3)\,\mu_B$ and $f=2200\text{K}/5\text{K}=440$ \cite{Koj97,Mot96}. For the frustrated zigzag chain compound SrCuO$_2$ (albeit with a simpler frustration than (NO)[Cu(NO$_3$)$_3$]) the magnetic order is incommensurate and even more suppressed ($m=0.033(7)\,\mu_B$ and $f=2600\text{K}/5\text{K}=520$) although the interchain exchange is larger ($J'/J\approx 0.1$) \cite{Zal99,Zal04}. Comparison to these examples suggests that the interchain exchange interactions in (NO)[Cu(NO$_3$)$_3$] are very weak and/or frustrated.

\begin{figure}
\includegraphics[width=0.9\columnwidth]{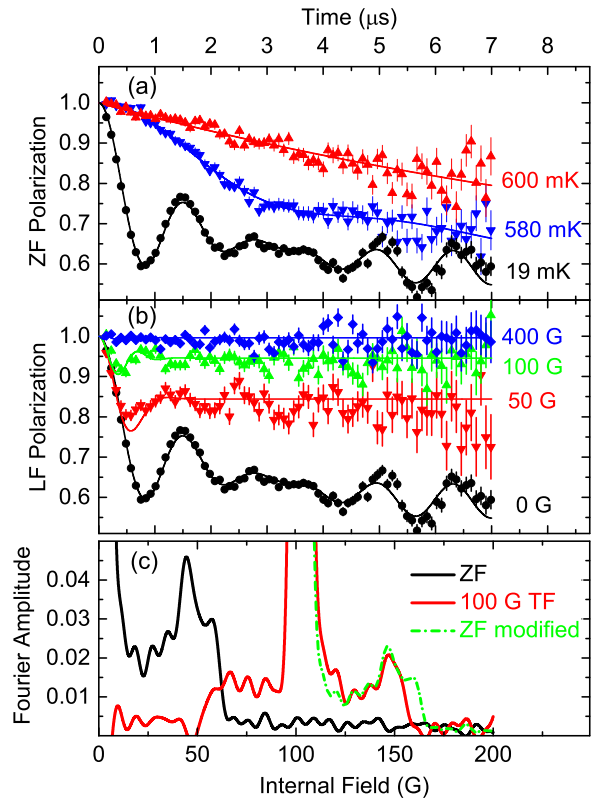}
\caption{(a) Zero-field $\mu$SR spectra. (b) Longitudinal field $\mu$SR spectra at 19\,mK. The lines are fits described in the text. (c) Field distribution at 19\,mK given in ZF and a TF of 100\,G. For comparison the ZF dataset is also plotted with a shift of 100\,G and divided by a factor of 2.}\label{Fig:muon}
\end{figure}

Zero field (ZF)-$\mu$SR spectra for selected temperatures are shown in Fig.~\ref{Fig:muon}a. Above the phase transition the spectra are characterized by weak exponential relaxation, typical for the presence of only diluted nuclear moments. At 580\,mK the influence of electronic moments is revealed by increased relaxation, while oscillations due to spontaneous muon spin precession are observed below this temperature. The observation of spontaneous muon spin precession indicates the presence of long-range magnetic order which produces an internal magnetic field at the muon site. From the beating pattern of the oscillations it is clear that there are at least two dominant fields. A single smooth field distribution could not produce this beating.

Below $T_N$ the data had to be fitted by the sum of two signals (in addition to a non-relaxing signal from the Ag sample holder)
\begin{align*}
P(t) &= g(T) P_{LRO}(t) + (1-g(T)) P_{SRO}(t)\\ 
&P_{LRO}(t) = \sum\limits_{n=1}^{2} f(n) e^{-0.5 \sigma(n)^2t^2} J_0(\gamma_\mu B_{max}(n)t)\nonumber\\ 
&P_{SRO}(t) = e^{-0.5\sigma_{SRO}^2t^2}.\nonumber
\end{align*}
Here, $g(T)$ gives the fraction of the precessing signal due to long-range magnetic order (LRO), while $(1-g(T))$ gives the relaxing signal due to disorder/short range order (SRO). $J_0$ is a zero-order Bessel function, $\gamma_\mu$ is the gyromagnetic ratio of the muon and $B_{max}$ is the maximum of the field distribution at the muon site \cite{Yao11}. Both LRO and SRO were simultaneously fitted by allowing the LRO fraction $g(T)$ to increase below $T_N$. The SRO was modeled with a Gaussian relaxation function and accounts for $\sim$30\,\% of the signal at the lowest temperature. It indicates a broad distribution of local fields possibly due to frustration, grain boundaries or magnetic domain walls although its origin in (NO)[Cu(NO$_3$)$_3$] is unknown. The LRO could be fitted at all temperatures by a sum of two Bessel functions with different internal field distributions indicating the presence of two inequivalent muon sites in the crystallographic lattice. These two sites were found to contribute equally $(f(1)=f(2))$ and to have exactly the same temperature dependence indicating that they observe the magnetic order from two different perspectives. The ratio between the two fields was $B_{max2} =1.23\,B_{max1}$ at all temperatures.

\begin{figure}
\includegraphics[width=0.7\columnwidth]{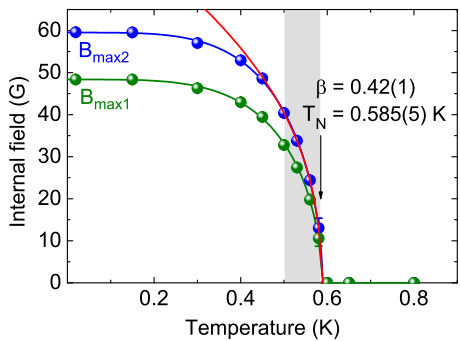}
\caption{Magnetic order parameter: Temperature dependence of the local magnetic fields $B_{max1}$ and $B_{max2}$, which are proportional to the ordered magnetic moment. The red line is a fit to $B_{\text{max2}}=B_0(1-T/T_N)^{\beta}$ for T$\geq$0.5\,K (shaded region).}\label{Fig:phase_diagram}
\end{figure}

The temperature dependence of the maximum internal magnetic field at both muon sites is shown in Fig.~\ref{Fig:phase_diagram}. Fitting both fields to the power law $B=B_0(1-T/T_N)^{\beta}$ for $T\geq$0.5\,K gives the Néel temperature, $T_N=585(5)$\,mK, in good agreement with the value of $T_N=580(10)$\,mK from specific heat.

The field distribution in the ordered state at 19\,mK is given by the real part of the Fast Fourier transform of the spectra (black curve in Fig.~\ref{Fig:muon}c). It is broad and asymmetric with two strong peaks at 44\,G and 57\,G and a pronounced tail to small fields. The weak internal fields at the muon sites suggests a small ordered magnetic moment in qualitative agreement with the neutron diffraction, although it should be emphasized that the muon sites are unknown. The field distribution is reminiscent of an Overhauser distribution which is typical of incommensurate magnetic order \cite{Yao11}. This is an important result because incommensurate magnetism usually arises from frustrated interactions. Since the order along the chains is commensurate as revealed by the INS, the order perpendicular to the chains must be incommensurate. This implies that the interchain interactions are frustrated.

The effect of a transverse magnetic field (TF) on (NO)[Cu(NO$_3$)$_3$] was also investigated. Fig.\,\ref{Fig:muon}c shows the field distributions at the two muon sites extracted from the TF-$\mu$SR at base temperature. A TF of 100\,G was applied vertically i.e. parallel to the $b$ axis. For comparison the external field is added to the ZF distribution and the amplitude is divided by 2 to correct for the summation of positive and negative field contributions which happens automatically in a ZF experiment. After correction, the distribution of the ZF measurement is surprisingly similar to the TF distribution. The peak describing the maximum field of muon site 1 appears at the same position (144\,G) with almost the same intensity. The external field simply shifts it leaving its shape unaffected, indicating that the internal fields at this site are all parallel to the external field, i.e. pointing along the $b$ axis. The peak describing the maximum field of muon site 2 shifts to lower fields in the TF data and is not visible anymore suggesting a substantial angle between the internal and external fields at this muon site.
 
The spatially constant spin direction found at muon site 1 implies that the magnetic order must be collinear. Together with the result that the magnetic order is incommensurate this suggests a sinusoidally modulated magnetic structure and rules out spiral order. In a spiral structure the fields at the crystallographically equivalent muon sites point in different directions and an applied TF would change the shape of the field distribution. A sinusoidal rather than spiral structure in a frustrated magnet may indicate anisotropic magnetism.

Longitudinal (LF)-$\mu$SR measurements were also performed to check for magnetic dynamics at and below $T_N$. In the presence of static magnetism, a magnetic field applied longitudinal to the initial muon spin direction shifts the polarization function to higher values \cite{Yao11}. At fields $\sim$10 times the internal field the muon spin is fully polarized and constant as a function of time with complete loss of oscillations. If the magnetic system were dynamic it would be impossible to fully regain the polarization since spin flip processes would continue to depolarize the muon spins which would be observed as a slow decay of the LF-$\mu$SR spectra.

The LF-$\mu$SR spectra at 19\,mK reveal the expected behavior for static magnetism (Fig.~\ref{Fig:muon}b). The lines are fits to a static Gaussian-Kubo-Toyabe function \cite{Yao11}. The polarization is almost constant for a LF of 400\,G. Thus 400\,G is sufficient to decouple the muons from the internal fields implying that the transverse components of the internal fields are $\sim$40\,G (consistent with the ZF results). The LF-$\mu$SR measurements at $T_N$ also reveal completely static magnetism. Since quantum fluctuations reducing the ordered moment must be present, they must be too fast for the muon time window which is $\sim$MHz. This is reasonable given the large intrachain exchange constant $J=142\,\text{K}=2.96$\,THz.


To summarize, inelastic neutron scattering shows by the presence of a spinon continuum that (NO)[Cu(NO$_3$)$_3$] is a 1D S-\nicefrac{1}{2} HAFC with $J=142(3)$\,K. Long-range magnetic order occurs at the highly reduced Néel temperature $T_N=585(5)$\,mK. The large ratio $|J|/T_N=243$ reveals strong suppression of magnetic order. Furthermore, the $C_p$ and $\mu$SR imply a small ordered moment while neutron diffraction gives an upper limit of $m\leq0.01\,\mu_B$. Evidence that the interchain interactions are competing comes from $\mu$SR, which shows that the magnetic order is an incommensurate spin density wave. Since the INS reveals commensurate magnetism along the chain, the order must be incommensurate perpendicular to the chains. We therefore conclude that (NO)[Cu(NO$_3$)$_3$] is a highly one-dimensional chain with frustrated interchain interactions. Thus this compound can be described by the Nersesyan-Tsvelik model with finite and competing values of $J'$ and $J_2$ although the ratio of these interactions and the proximity of the system to the special point $J'=2J_2$ is unknown.

\begin{acknowledgments}
We thank E. Shimshoni for introducing us to the Nersesyan-Tsvelik model and setting up this collaboration. We also thank K. Siemensmeyer, B. Klemke and K. Prokes for their help with the E4 neutron diffraction measurement. This work was partially supported by RFRB grants 13-02-00174, 14-02-00111, 14-02-92002. OSV acknowledges the grant of RF President MK-7138.2013.2.
\end{acknowledgments}

\end{document}